\documentclass[pra,aps,twocolumn,10pt,showpacs,groupedaddress,superscriptaddress,floatfix,notitlepage]{revtex4-2}
\usepackage{amsmath, physics, amsfonts,amssymb,graphics,graphicx,epsfig,color,times}
\usepackage[utf8x]{inputenc}
\usepackage{color}
\usepackage{bbm, dsfont} 
\usepackage{subfigure}
\usepackage{hyperref}
\usepackage{mathrsfs}
\usepackage{verbatim}
\begin{document}
\title{Interpretable machine-learning identification of the crossover from subradiance to superradiance in an atomic array}
\author{C. Y. Lin}
\email{jacky0921199997@gmail.com}
\affiliation{Institute of Atomic and Molecular Sciences, Academia Sinica, Taipei 10617, Taiwan}
\author{H. H. Jen}
\email{sappyjen@gmail.com}
\affiliation{Institute of Atomic and Molecular Sciences, Academia Sinica, Taipei 10617, Taiwan}

\date{\today}
\renewcommand{\r}{\mathbf{r}}
\newcommand{\f}{\mathbf{f}}
\renewcommand{\k}{\mathbf{k}}
\def\p{\mathbf{p}}
\def\q{\mathbf{q}}
\def\bea{\begin{eqnarray}}
\def\eea{\end{eqnarray}}
\def\ba{\begin{array}}
\def\ea{\end{array}}
\def\bdm{\begin{displaymath}}
\def\edm{\end{displaymath}}
\def\red{\color{red}}
\pacs{}
\begin{abstract}
Light-matter interacting quantum systems manifest strong correlations that lead to distinct cooperative spontaneous emissions of subradiance or superradiance. To demonstrate the essence of finite-range correlations in such systems, we consider an atomic array under the resonant dipole-dipole interactions (RDDI) and apply an interpretable machine learning with the integrated gradients to identify the crossover between the subradiant and superradiant sectors. The machine shows that the next nearest-neighbor couplings in RDDI play as much as the roles of nearest-neighbor ones in determining the whole eigenspectrum within the training sets. Our results present the advantage of machine learning approach with explainable ability to reveal the underlying mechanism of correlations in quantum optical systems, which can be potentially applied to investigate many other strongly interacting quantum many-body systems.             
\end{abstract}
\maketitle
\section{Introduction}
Resonant dipole-dipole interactions (RDDI) \cite{Stephen1964, Lehmberg1970} arise in light-matter interacting systems owing to multiple scatterings or reabsorption of light in the medium. This collective and pairwise interaction results in the subradiance or superradiance \cite{Dicke1954, Gross1982}, which respectively corresponds to a longer or shorter lifetime of cooperative spontaneous emissions \cite{Pellegrino2014, Scully2015, Plankensteiner2015, Jennewein2016, Guerin2016, Bromley2016, Jen2016_SR, Sutherland2016, Bettles2016, Shahmoon2017, Jen2017_MP, Garcia2017, Plankensteiner2017, Jenkins2017, Bhatti2018, Guimond2019} compared to the natural decay. Their time scales can span wide ranges across the intrinsic lifetime of individual particles, and the range becomes wider in scales especially for a dense medium \cite{Jen2016_SR}. As the number of quantum emitters in a dense medium grows, the complexity of RDDI escalates, and the system dynamics becomes intricate owing to its emerging multiple scales in space and time. There is no easy way to characterize the system's radiation property under the RDDI \cite{Chomaz2012} without resolving its full eigenspectrum. This is due to the long-range nature of RDDI, which makes a truncation of interaction range deficient to genuinely describe the system's dynamics, despite of a recent effort of using renormalization group method to reduce a system under strong RDDI to an inhomogeneously broadened ensemble of weakly interacting atoms \cite{Andreoli2021}.

Recently, machine learning (ML) \cite{Goodfellow2016, Mehta2019} has been widely applied in many versatile physical domains \cite{Carleo2019}, which shows the power in identifying topological phases of matter \cite{Carrasquilla2017, Pilozzi2018, Rodriguez-Nieva2019} or classifying the experimental data from two-dimensional Fermi-Hubbard models \cite{Bohrdt2019}, finding improved solutions to long-distance quantum communication \cite{Wallnofer2020} and revealing the arrow of time in thermodynamics \cite{Seif2021}. Aside from utilizing ML in physical sciences, modern ML-based techniques have been applied in many areas of industry, and shown the ability of pattern recognition and the advantage of decision making \cite{Carleo2019}, for example.  

In contrast to the common emphasis on the prediction power of ML in fundamental sciences and modern technologies, here we present an interpretable ML \cite{Molnar2020, Adadi2018} that directly analyzes the interaction components and allows causal explanations in our model system of an atomic array under RDDI. The causal explanation relies on the attribution images in an interpretable ML, where we can extract their importance features that lead the trained machine to the prediction of the labels or the categories the images belong to. We use the attributions from the integrated gradients on the trained machine, where the essential crossover feature from the subradiant to superradiant eigen-decay rates can be revealed in either strong or weak RDDI. We find that the next nearest-neighbor (NN) couplings in RDDI are as crucial as the NN ones, which indicates the indispensable role of finite-range interaction in the many-atom system under RDDI. The interpretable ML can serve as a complimentary approach to our understanding of the light-matter interacting systems, and the explainable attribution we obtain here can be applied further to strongly correlated condensed-matter systems to uncover distinct features of spatial or time correlations.   

In this article, we consider a system of singly-excited atomic array which mediates RDDI and introduce its long-range interaction kernel in Sec. II. In Sec. III, we discuss how we implement the machine learning approach in the training and testing stages to predict the eigen-decay rates of the atomic array. In Sec. IV, we introduce the attributions from the integrated gradients as an interpretable machine learning, which can be used to reveal the importance features in the subradiant and superradiant sectors of the spectrum and identify the crossover between them. The results also show the essential roles of finite-range interactions beyond the nearest-neighbor ones. Finally we discuss and conclude in Sec. V.  

\section{Resonant dipole-dipole interactions}
The pairwise RDDI emerge from a reservoir of quantized bosonic fields interacting with the atoms \cite{Lehmberg1970}. After the reservoir degrees of freedom are traced out, the Heisenberg equation of arbitrary quantum operator $\hat{Q}$ can be expressed in Lindblad forms as   
\bea
\frac{d\hat{Q}}{dt} =&& -i\sum_{\mu\neq\nu}^N\sum_{\nu=1}^N G_{\mu\nu}[\hat{Q},\hat{\sigma}_\mu^+\hat{\sigma}_\nu^-] \nonumber\\
&&-\sum_{\mu=1}^N\sum_{\nu=1}^N\frac{F_{\mu\nu}}{2}\left(\hat{\sigma}_\mu^+\hat{\sigma}_\nu^-\hat{Q}+\hat{Q}\hat{\sigma}_\mu^+\hat{\sigma}_\nu^- -2\hat{\sigma}_\mu^+\hat{Q}\hat{\sigma}_\nu^-\right),\nonumber\\\label{Q}
\eea
where the dipole operator is $\hat{\sigma}_\mu^-$ $\equiv$ $|g\rangle_\mu\langle e|$, and $\hat{\sigma}_\mu^-$ $\equiv$ $(\hat{\sigma}_\mu^+)^\dag$ for two-level quantum emitters with the ground and the excited states as $|g\rangle$ and $|e\rangle$, respectively. The collective and pairwise frequency shifts $G_{\mu\nu}$ and decay rates $F_{\mu\nu}$ are 
\bea
F_{\mu\nu}(\xi)\equiv&&
\frac{3\Gamma}{2}\bigg\{\left[1-(\hat\p\cdot\hat{r}_{\mu\nu})^2\right]\frac{\sin\xi}{\xi}\nonumber\\
&&+\left[1-3(\hat\p\cdot\hat{r}_{\mu\nu})^2\right]\left(\frac{\cos\xi}{\xi^2}-\frac{\sin\xi}{\xi^3}\right)\bigg\},\label{F}\\
G_{\mu\nu}(\xi)\equiv&&\frac{3\Gamma}{4}\bigg\{-\left[1-(\hat\p\cdot\hat{r}_{\mu\nu})^2\right]\frac{\cos\xi}{\xi}\nonumber\\
&&+\left[1-3(\hat\p\cdot\hat{r}_{\mu\nu})^2\right] \left(\frac{\sin\xi}{\xi^2}+\frac{\cos\xi}{\xi^3}\right)\bigg\}\label{G}, 
\eea
where $\Gamma$ is the intrinsic decay rate of the excited state, dimensionless $\xi$ $=$ $|\k_L| r_{\mu\nu}$ with the wave vector $|\k_L|=2\pi/\lambda$, transition wavelength $\lambda$, interparticle distance $r_{\mu\nu}$ $=$ $|\mathbf{r}_\mu-\mathbf{r}_\nu|$ $=$ $d_s|\mu-\nu|$ and dipole orientation $\hat\p$. We use $d_s$ to quantify the strength of RDDI in the periodic array and assume $\hat \p\perp\hat r_{\mu\nu}$ throughout the paper. The above forms of RDDI look exactly as the interaction between two permanent dipoles but are induced by light excitations by nature.   

\begin{figure}[t]
\centering
\includegraphics[width=8.5cm,height=5.5cm]{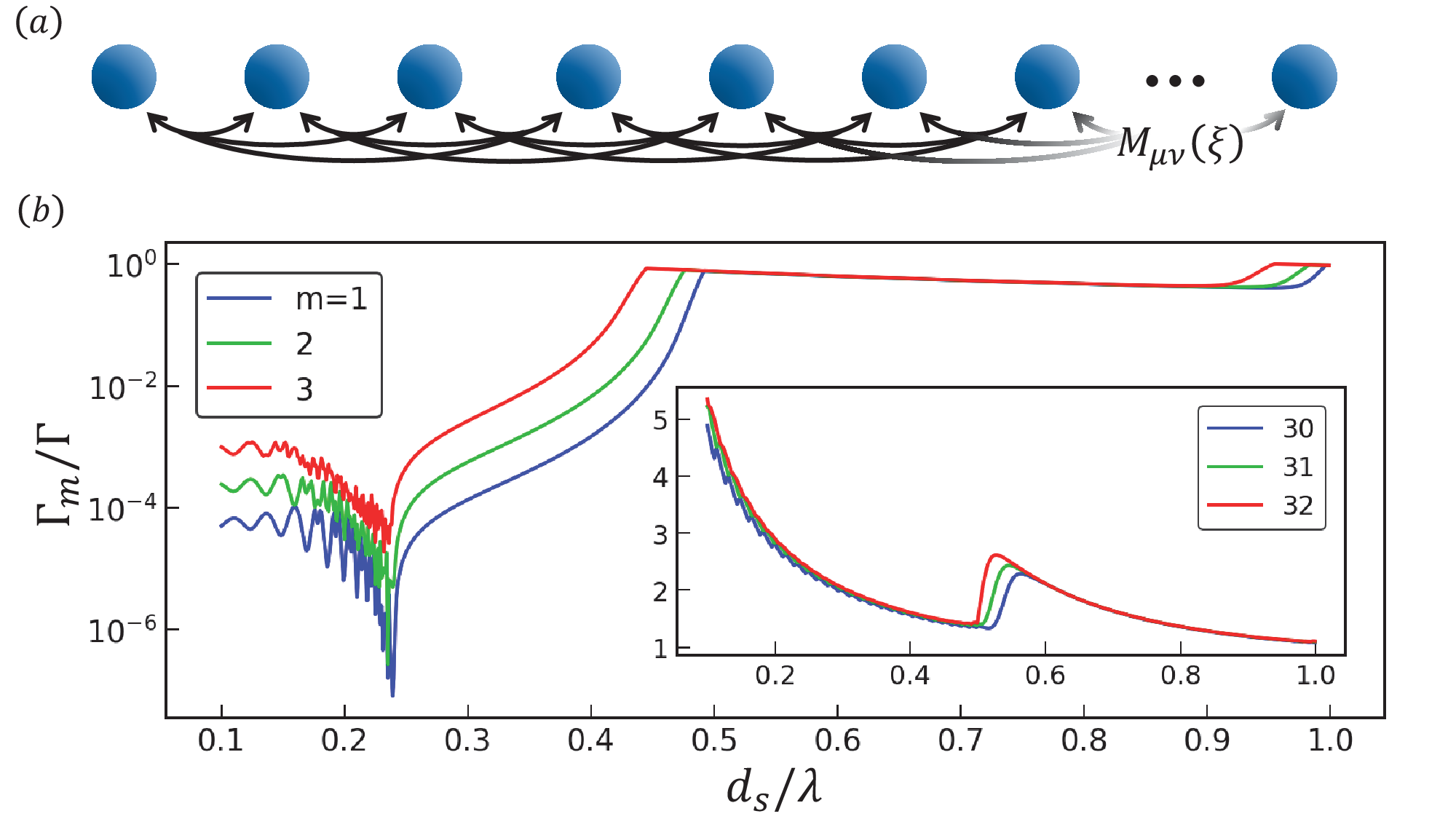}
\caption{Schematic atomic array under RDDI and eigen-decay rates for subradiance and superradiance. (a) A system of atomic array with equidistant interparticle separations $\xi/|\k_L|$ under RDDI has pairwise interaction forms denoted by $M_{\mu\nu}(\xi)$. (b) Eigen-decay rates $\Gamma_m$ for $N=32$ are obtained in an ascending order, where the first three subradiant and the last three superradiant eigenvalues are shown for various distances $d_s$.}\label{fig1}
\end{figure}

When we consider single excitation in an atomic array as shown in Fig. \ref{fig1}(a), the complete Hilbert space in bare states $|\phi_\mu\rangle$ $=$ $\hat\sigma_\mu^\dag |g\rangle^{\otimes N}$ along with the probability amplitudes $b_\mu(t)$ are sufficient to describe the system's dynamics. Within this subspace, we can obtain the system's time evolution $|\Phi(t)\rangle$ $=$ $\sum_{\mu=1}^N b_\mu(t)|\phi_\mu\rangle$ from Schr\"{o}dinger's equation,  
\bea
i\frac{\partial}{\partial t}|\Phi(t)\rangle=\tilde H|\Phi(t)\rangle,\label{nonH}
\eea  
where a non-Hermitian Hamiltonian $\tilde H$ reads 
\bea
\tilde H &=&-i\sum_{\mu=1}^N\sum_{\nu=1}^N M_{\mu\nu} \hat\sigma_\mu^\dag\hat\sigma_\nu, 
\eea
and the elements of the associated interaction matrix $M$ becomes
\bea
M_{\mu\nu}\equiv -\frac{F_{\mu\nu}}{2}+iG_{\mu\nu}\delta_{\mu\neq\nu}. 
\eea
The $M$ in general is a symmetric matrix owing to the symmetries $F_{\mu\nu}=F_{\nu\mu}$ and $G_{\mu\nu}=G_{\nu\mu}$, which can be diagonalized by similarity transformations. We note that this kind of singly-excited coupled dipoles can be simulated by classical analogs of dipole oscillators, which results in a classical many-body interacting system.  

In Fig. \ref{fig1}, we take the example of an atomic array with $N$ $=$ $32$ and obtain the eigen-decay rates of $\Gamma_m$ $=$ $-2$Re$(\lambda_m)$ in an ascending order, which is numerically calculated by diagonalizing $M$ with the eigenvalues $\lambda_m$. The subradiant and superradiant decay rates significantly deviate from the noninteracting regime ($\Gamma_m$ $\sim$ $\Gamma$) when $d_s/\lambda$ $\lesssim$ $0.5$. In this strong interacting regime, contrasted scales of decay rates that span wide orders of magnitude can be seen from the most subradiant to the superradiant ones. This shows the richness of these collective spontaneous emissions from many quantum emitters, where the superradiant emission can be described by a uniformly-distributed singly-excited symmetric state \cite{Scully2006, Eberly2006, Mazets2007}, whereas the subradiance can be formed from a more subtly-prepared phase-imprinted Hilbert space \cite{Jen2016_SR, Jen2017_MP, Jen2018_SR1, Jen2018_SR2}.  

It seems that it is inevitable to take a full account of the eigenspectrum of quantum systems to genuinely describe its dynamics, and the faithfulness of this approach can be easily broken if a truncation of interaction ranges is applied or only near-field or far-field RDDI is taken into account \cite{Andreoli2021}. It is this effective or renormalized perspective of RDDI that underestimates the importance of atom-atom correlations hidden in the interacting system. Below we employ a machine learning approach to study the system of Fig. \ref{fig1}(a) and try to answer the question of how crucial the range of RDDI should be in determining the subradiant and superradiant phenomena of light.   

\section{Machine learning the eigenspectrum}

\begin{figure}[t]
\centering
\includegraphics[width=8.5cm,height=8cm]{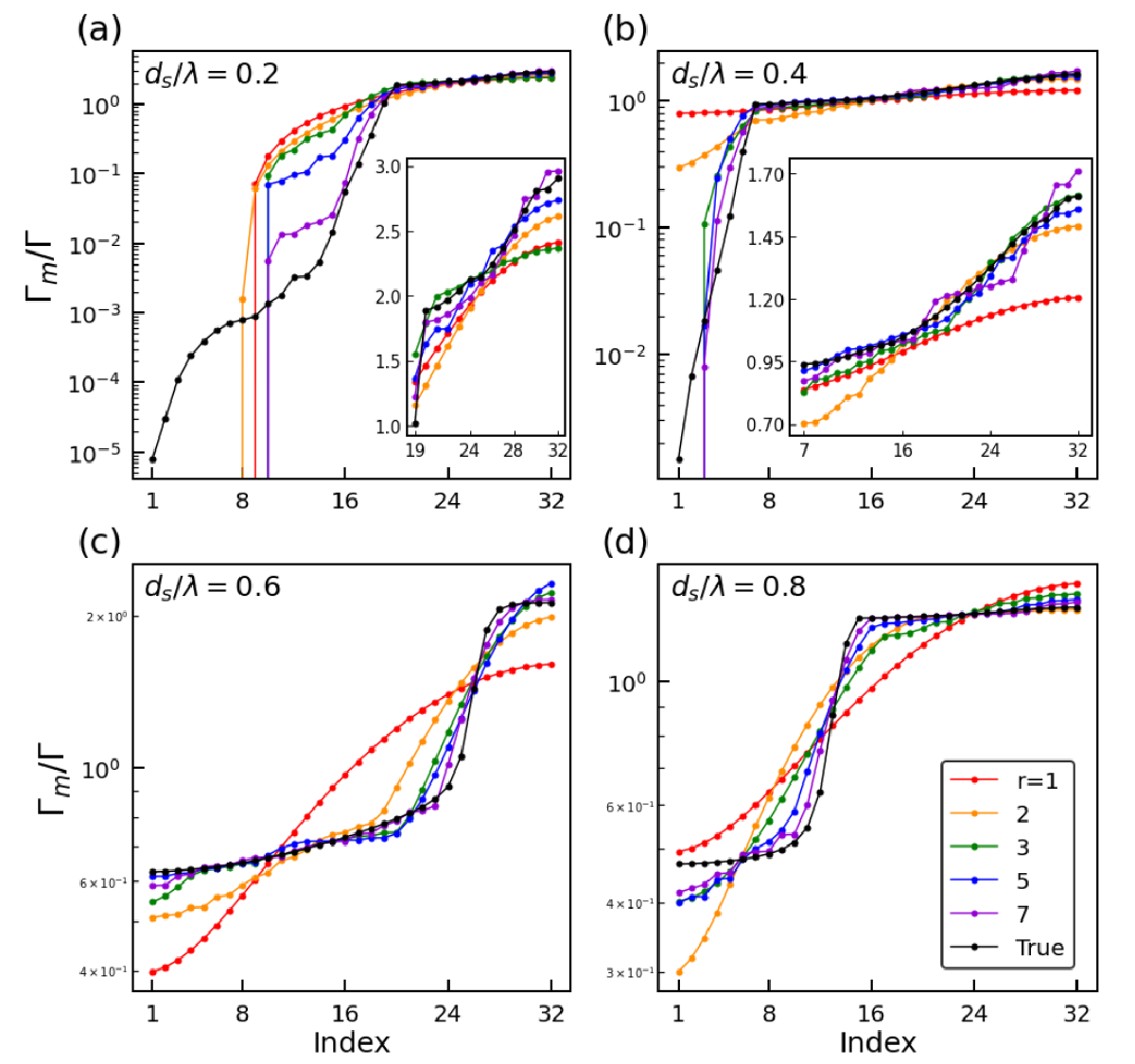}
\caption{Eigen-decay rates $\Gamma_m$ from various truncation ranges $r$ ($r=1$ and $2$ for NN and next NN coupling ranges) for $N=32$. In a direct diagonalization, we compare the eigen-decay rates in logarithmic scales in an ascending order for different $r$ at $d_s/\lambda$ $=$ $0.2$, $0.4$, $0.6$, and $0.8$, respectively in (a-d). As $r$ increases, $\Gamma_m$ approaches the genuine one with all coupling ranges (up to $r=7$). The insets in (a) and (b) are zoomed in of the superradiant sectors, and some of the subradiant eigenvalues in (a) and (b) become negative and unphysical (unable to show in logarithmic scales) owing to the brute-force truncations.}\label{fig2_tr}
\end{figure}

It is well-known that the near-field terms of Eqs. (\ref{F}) and (\ref{G}) are more significant than the far-field ones. This is evident between any two atoms, while it is not obvious that when more atoms are involved, to what extent of RDDI ranges or which interaction component matters the most to attribute to the collective decay rates of the eigenspectrum, since the corresponding eigenvectors generally involve all atoms to all ranges. It seems that we can truncate the interaction ranges in RDDI kernel and use a direct diagonalization method to demonstrate how the long-range interaction modifies the eigenspectrum. In Fig. \ref{fig2_tr}, we compare the cases with various truncated ranges $r$ and find that they approach the genuine case when $r$ increases. We can see that the true eigen-decay rates require all-range couplings in the RDDI kernel. Whether a certain range of couplings is crucial or not in determining the eigenvalues might be revealed by designing a couple of ways of removing or including the off-diagonal elements of RDDI. However, this does not provide a systematic way to study the essential or particular ranges of interactions that lead to the genuine eigenvalues. Furthermore, we note that the eigen-decay rates would give negative and unphysical values owing to the truncations in strongly interacting regimes as shown in Figs. \ref{fig2_tr}(a) and \ref{fig2_tr}(b).  

Here we would like to understand the importance feature of RDDI that leads to the collective eigen-decay rates by using the machine learning approach in a systematic way with attribution technique. We will show in the next section that the trained machine reaffirms the important near-field interactions relatively, and furthermore, it presents a crossover behavior from the subradiant to superradiant sectors of the eigenspectrum. This provides an identification between these sectors with significantly different time scales and also suggests an insufficient truncation of RDDI at the nearest-neighbor sites. 

\begin{figure}[tb]
\centering
\includegraphics[width=8.5cm,height=5.5cm]{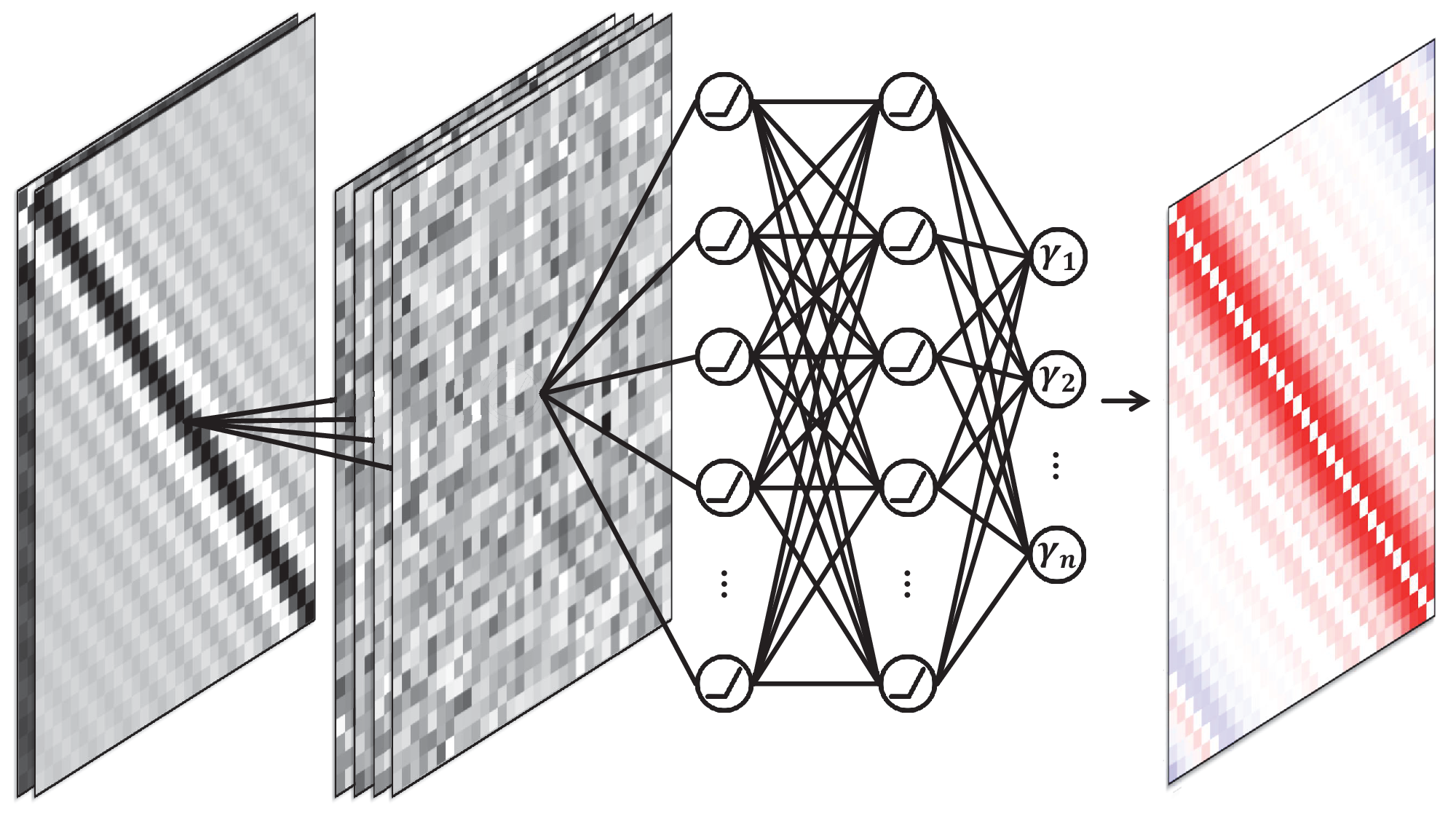}
\caption{Attribution calculation by integrated gradients after learning the eigen-decay rates from the interaction kernel $M$ with CNNs. Subject to piecewise-defined activation functions in neurons, the leftmost input images are two real-valued layers of the real and imaginary parts of $M$. After convolving the input images with a set of learnable filters, pixels of convolved images are fully connected to deep neural networks, which then predicts the eigen-decay rates. Using the method of integrated gradient, pixels in interaction kernel images attribute differently to predict the eigen-decay rates at a certain atomic spacing. Most attribution images show that the nearest neighbor and the next nearest neighbors are more critical to the collective radiations.}\label{fig2}
\end{figure}

Before introducing the attribution technique as an interpretable machine learning, we first train a convolution neural network (CNN) as shown in Fig. \ref{fig2}. A CNN usually consists of three types of layers: convolution layers, pooling layers and dense layers. Convolution layers convolve input images with image filters to extract possible hidden patterns. The size of filters is usually smaller than the size of the input images. Nonetheless, we find that when both sizes are equal, the images attribution images become symmetric naturally rather than post-symmetrizing them manually as required by physical insights. Pooling layers usually take either the maximum or the average value across convolved images. Here we remove this layer since every elements in our convolved images should participate in the collective effect we investigate in this work. Besides, the size of our interaction kernel images is not that large, and so we do not need the pooling layers to reduce the burden of training variables. The final dense layers are fully connected layers which could enhance the complexity and nonlinearity of the datasets by introducing nonlinear activation functions across the dense layers. ReLU, the nonlinear piecewise-defined activation function we used, cannot take in the complex numbers contained in an interaction kernel. Hence, we split the real and imaginary part of kernels and combined them as two-layer images.

We then use the CNN consisting of one convolution layer followed by two fully-connected dense layers to regenerate the eigen-decay rates in logarithmic scales from two-layered interaction kernel images which are composed respectively of the real and imaginary parts of RDDI. Instead of focusing on the prediction power of machine learning the physical observables or parameters, we blend the training and testing regions randomly and set $9:1$ as the train-to-test-ratio, where the data contains $300000$ sets of interaction kernels and we train their corresponding decay rates from $d_s/\lambda$ $=$ $0.1$ to $1.0$. The number of learnable filters in convolution layers is four times the size of atomic array, and the numbers of neurons in each dense layer are both two times the size of atomic array. The machine is expected to completely learn the relationship between the eigenspectrum and the patterns in the interaction kernel images. To determine the number of dense layers in CNN, we have also tried more than two of them, where we do not find significantly improved accuracies in testing. For the training-to-testing ratio, we note that a less training ratio would decrease the performance of machine, while on the other hand, a machine lacks the learning capability if the ratio is too high in the training stage. We fix the ratio as $9:1$ along with large enough datasets for the best of machine’s performance in testing within reasonable computation time. 

\begin{figure}[tb]
\centering
\includegraphics[width=8.5cm,height=5.5cm]{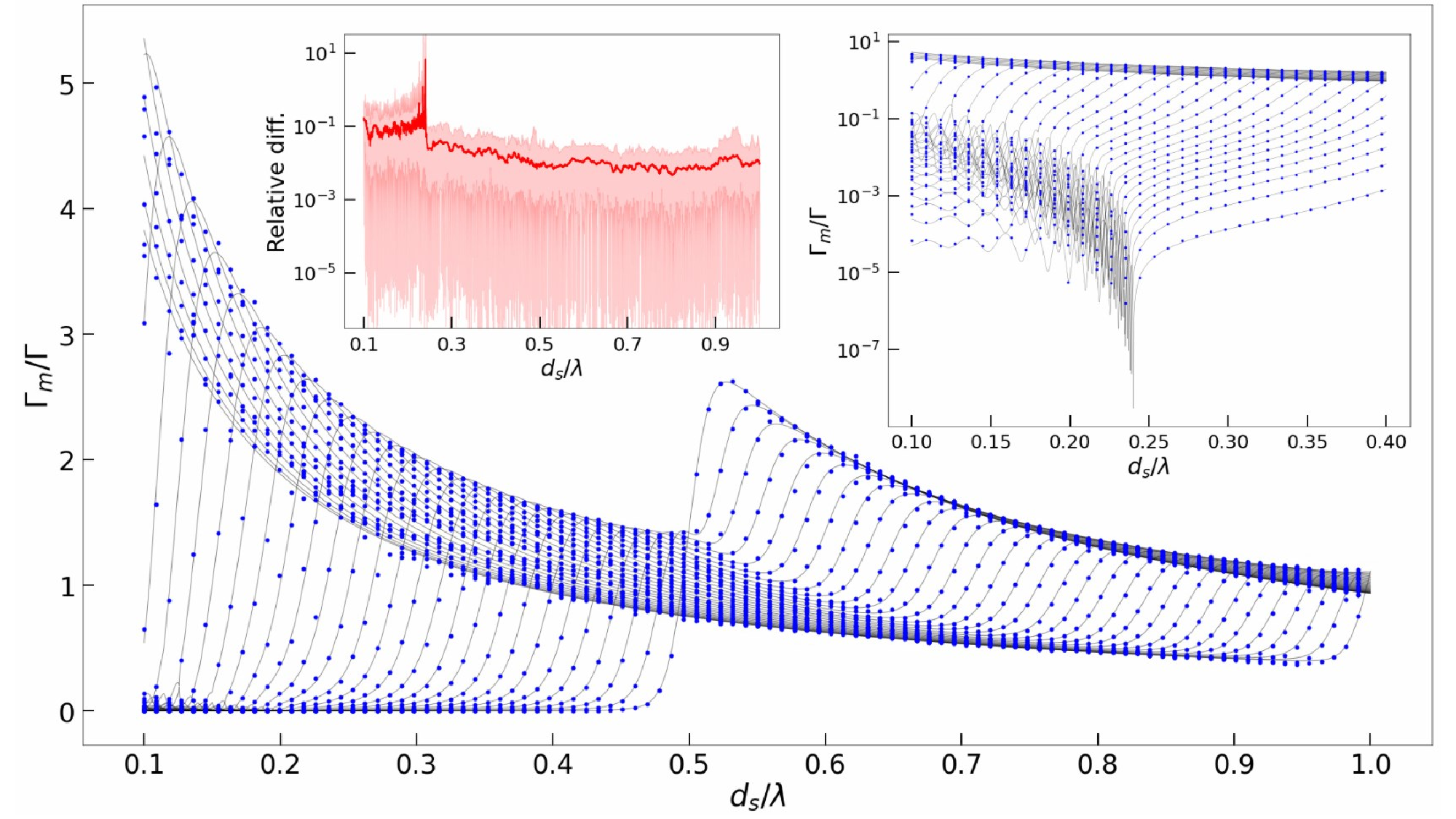}
\caption{Tests of the trained CNNs in a range of atomic spacing for the system size of $N$ $=$ $32$. For every atomic spacing, the CNNs predict $32$ eigen-decay rates in an ascending order denoted as blue dots on top of the true ones in black solid lines. The right inset zooms in the prediction of the subradiant decay rates at small interparticle distances. In this regime, the prediction accuracy can be improved by taking the logarithm of the eigen-decay rates, but the accuracy is limited because the fast-oscillating parts are relatively hard to learn. The relative differences (diff.) and their mean value (solid red) are plotted in the left inset. Most of them are within a level of $10^{-2}$ for $d_s/\lambda\gtrsim 0.3$, and the accuracy drops at $d_s/\lambda\lesssim 0.25$ for some of the most subradiant modes.}\label{fig3}
\end{figure}

To demonstrate the performance of machine, in Fig. \ref{fig3}, we show the comparison between the predicted eigen-decay rates from the trained machine and the true ones directly from diagonalization. The machine can mostly predict the eigen-decay rates well, which makes the interpretable attributions of images we calculate later more convincing. A relatively harder learning parameter regime appears in the subradiant sectors of eigen-decay rates at small atomic spacings, which is shown in the inset of Fig. \ref{fig3}. In this limited region, the prediction accuracy drops a little owing to the abrupt changes of the eigenvalues. We note that the eigenvalues can be easily calculated numerically on a regular computer, and therefore, we do not focus on the capability or efficiency of the predictability of the trained machine in the eigenspectrum. The testing results in Fig. \ref{fig3} intend to show that the trained machine can provide useful and informative attributions we obtain later in the next section. In addition, we can also identify the confidence level of the trained machine in various parameter regimes of interparticle distances we focus here. 

\section{Crossover in the images of attribution}

Here we introduce the integrated gradients \cite{Sundararajan2017} as an interpretable machine learning which intends to explain the relationship between the features of input images and the predicted labels from trained machines. The input images are the exact RDDI kernels that are used for training and testing in the CNNs of the machine learning, and the output gives the attribution images from the integrated gradients. What is explained or extracted in the attribution images at the output is the associated importance feature that leads to the predicted labels which are the eigen-decay rates here. Applying local gradients on the original input images relative to the predicted labels is usually an intuitive way to assign features that the trained machines `see'. However, the local gradients would approach to zero when a model has completely learned a certain feature connected to a label, since it does not affect the predictions by making a small deviation of pixels on the original images. As steep gradients may occur at any interpolated images from a baseline image to an original one, the integrated gradients technique includes all gradients along them and thus can reflect more distinctive features. This can be achieved since integrated gradients comply with the properties of sensitivity and the invariant implementation, where identical attributions should be obtained in two equivalent networks \cite{Sundararajan2017}. 

We take the image recognition as a demonstration of attribution technique in the reference of \cite{Sundararajan2017}. The input image is recognized and labeled as a fireboat, where the image involves a bridge above several water sprays in different directions from a fireboat on a river. The image is taken along the river level and its background is daytime sky. The simple gradient between the input and the baseline images shows a blurred output image mixing the water sprays and the sky, whereas the integrated gradients image presents a clear feature of water sprays above the fireboat, indicating the important elements of the input image that lead to the trained machine’s predicted label, a fireboat here.

The integrated gradients are defined as \cite{Sundararajan2017}
\bea
{\rm IntegratedGrads}_i^{\rm approx} (x)=&&\sum_{k=1}^m \frac{\partial F(x'+k/m\times (x-x'))}{\partial x_i}\nonumber\\&&\times \frac{(x_i-x_i')}{m}.     
\eea 
This represents the integrated gradients algorithm with $m$-steps Riemann approximation of the integral along the $i^{\rm th}$ dimension from an baseline image $x'$ to an input image $x$, where $\partial F(x)/(\partial x_i)$ is the gradient of a function $F$ which is defined as the negative squared difference between the predicted eigen-decay rate of the interpolated image and the one of the original image. This definition indicates the inclination to interpret what causes the machine make the final prediction. After the Riemann sum over the gradients of interpolated images $[x'+k/m\times (x-x')]$, the algorithm yields the attribution images. The baseline image, on which a starting point to apply gradients, is usually set to an image where no desired feature appears. Considering the feature of the collective effect we investigate here, we set the baseline image as a non-interacting kernel, where only the diagonal terms are nonzero as the natural decay rates, while the off-diagonal parts are vanishing owing to the weak RDDI when $d_s/\lambda$ $\gg$ $1$.

\begin{figure}[t]
\centering
\includegraphics[width=8.5cm,height=8.5cm]{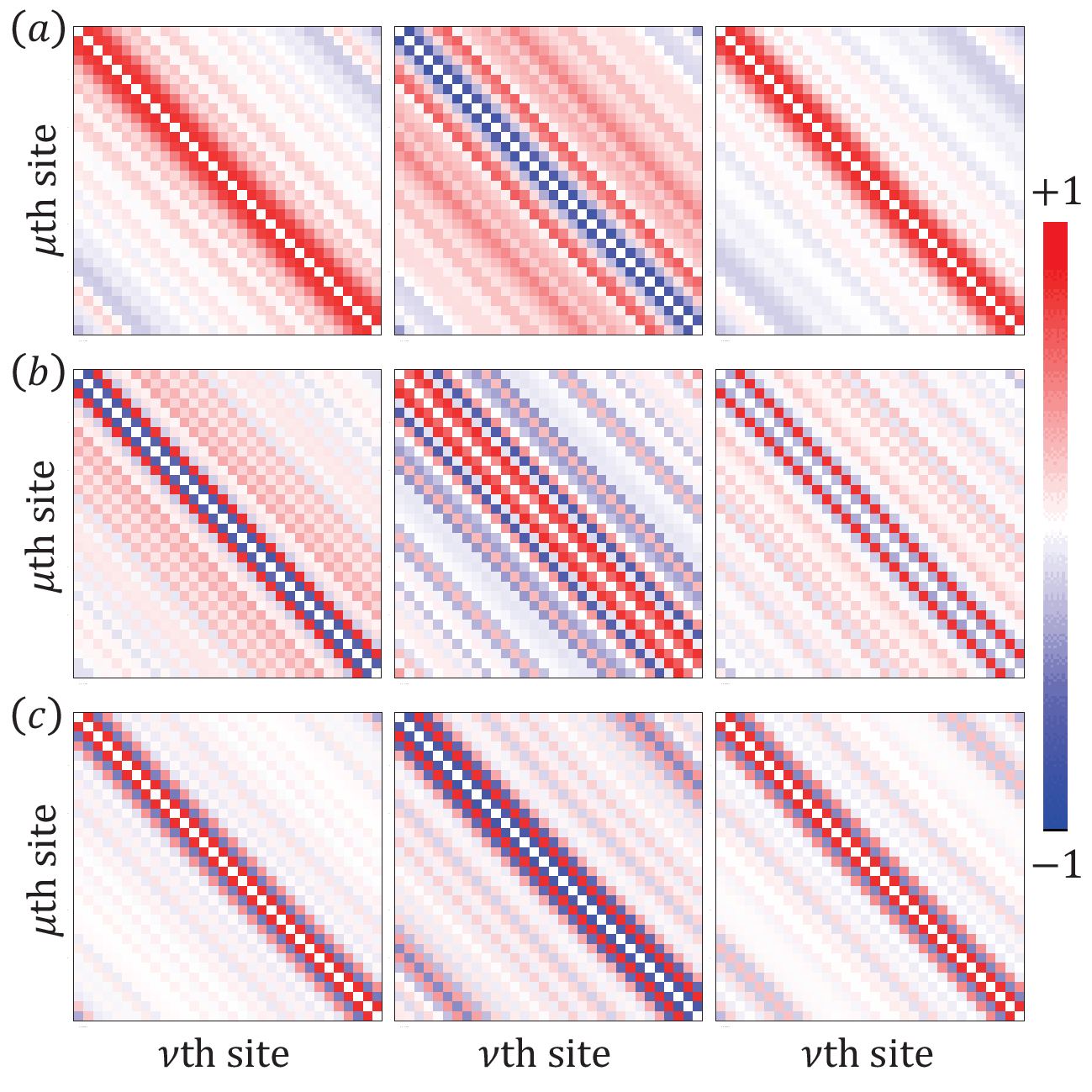}
\caption{Attribution images of crossover from subradiant to superradiant decay rates at $d_s/\lambda$ $=$ $0.25$, $0.5$ and $0.75$ in (a-c), respectively, for $N$ $=$ $32$. These colors of images are normalized to their own maximal and absolute values of pixels, and the faint color corresponds to near-zero values. The indices $m$ of the eigen-decay rates $\Gamma_m$ in (a-c) from left to right are $(1, 17, 32)$, $(1, 21, 32)$ and $(1, 15, 32)$, respectively. The attribution images of the most sub- and super-radiant decay rates (the first and the $N$th ones) present almost the same structure of importance in the nearest-neighbor and the next nearest-neighbor coupling ranges, which is universal for every atomic spacings. The attributions for the crossover regions of $\Gamma_m\sim\Gamma$ are distinguishable from both subradiant or superradiant regimes, which spread further into the off-diagonal parts.}\label{fig4}
\end{figure}

In Fig. \ref{fig4}, we use $m$ $=$ $1000$ steps in the integrated gradients and present the attribution images for three different interparticle distances from strong to moderate RDDI. Each pixels of the images are normalized within respective parameter regimes of interparticle distances and present the importance feature, where the scale close to one represents the most relevant and sensitive pixel that leads to machine's predictions of the eigen decay rates. The negative scale, on the other hand, means the counteractive effect on its predictions. The attribution images of the most subradiant and superradiant decay rates present similar feature importance, whereas the crossover regions of some representative eigen-decay rates $\Gamma_m\sim\Gamma$ distinguish from the subradiant and superradiant regimes (SSRs). For a smaller $d_s$ under strong RDDI in Fig. \ref{fig4}(a), both NN and next NN coupling ranges are important for SSRs, in huge contrast to the spreading features of the crossover regime in its off-diagonal parts along with a suppression at short distance. This indicates that the crossover region with the eigen-decay rates close to the intrinsic ones correspond to the interaction range at a longer distance. This makes sense since RDDI at longer distances have a weaker effect on the collective decay rates, leading to a flat spectrum of $\Gamma_m\approx\Gamma$. 

When we increase $d_s$ in Figs. \ref{fig4}(b) and \ref{fig4}(c), we find that the next NN and the NN coupling range stands out in the SSRs, respectively. By contrast, the respective crossover region has extending off-diagonal stripes of positive and negative features, which is most prominent in the middle plot of Fig. \ref{fig4}(b). For longer interparticle distances, long-range interferences of RDDI could arise and lead to relatively complex structures in the crossover regimes. In general, the attributions in SSRs manifest more significantly at short distances compared to the crossover regime, which allows us to identify the onset to move from the subradiance to superradiance. 

\begin{figure}[t]
\centering
\includegraphics[width=8.5cm,height=8.5cm]{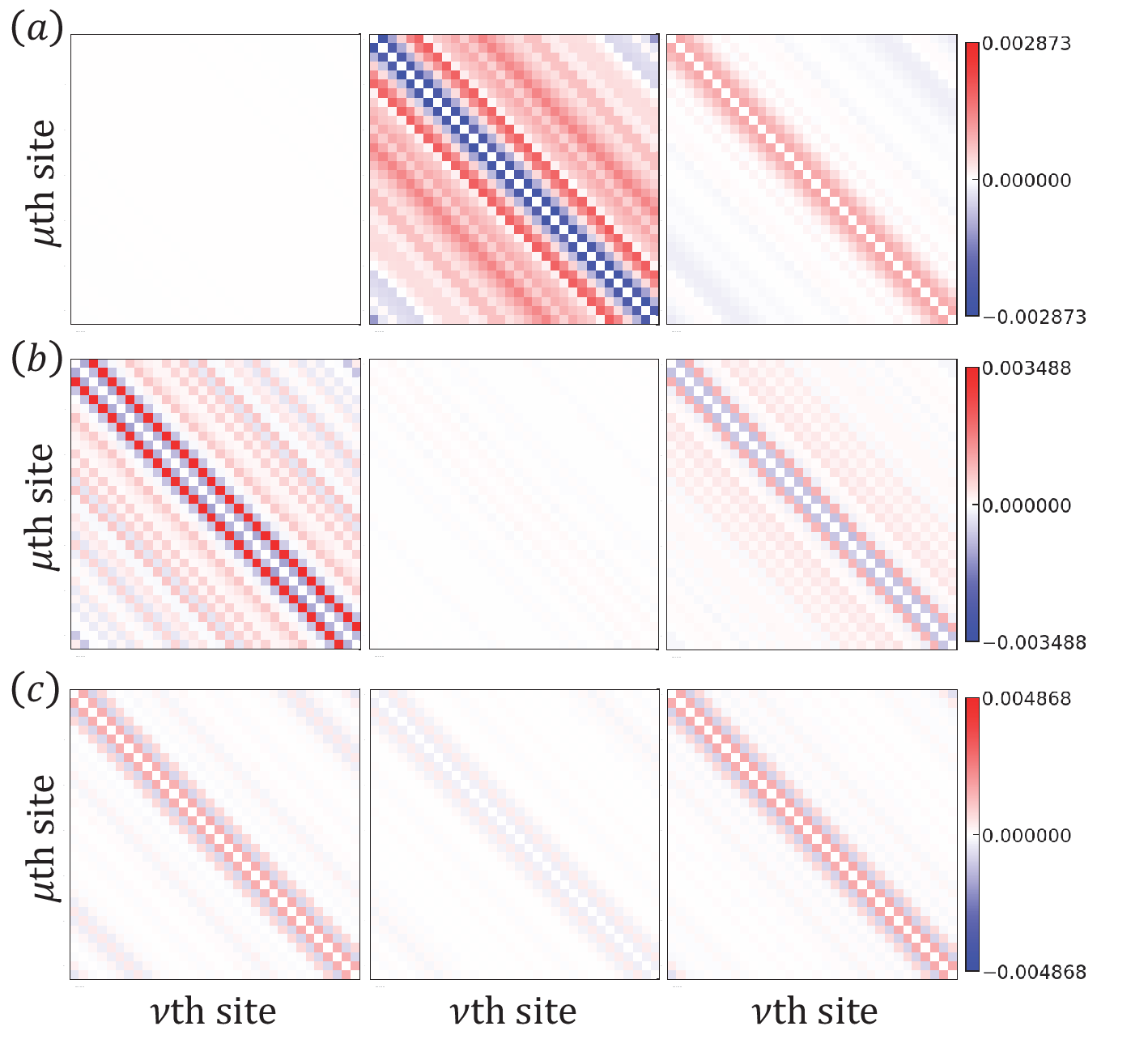}
\caption{Attribution images with the same conditions as in Fig. $5$ but with renormalized colors to the maximal absolute value of all attribution pixels at a certain atomic spacing. (a) The faint images take places in the subradiance, indicating the attributions from the corresponding interaction kernel images to the subradiance are relatively weak in the region of small atomic spacings. In (b) and (c), the relatively weak attributions from the interaction kernel images to the eigen decay rates appear in the crossover region from sub- to super-radiant decay rates at a larger atomic spacing.}\label{fig4_s1}
\end{figure}

\begin{figure}[t]
\centering
\includegraphics[width=8.5cm,height=8.5cm]{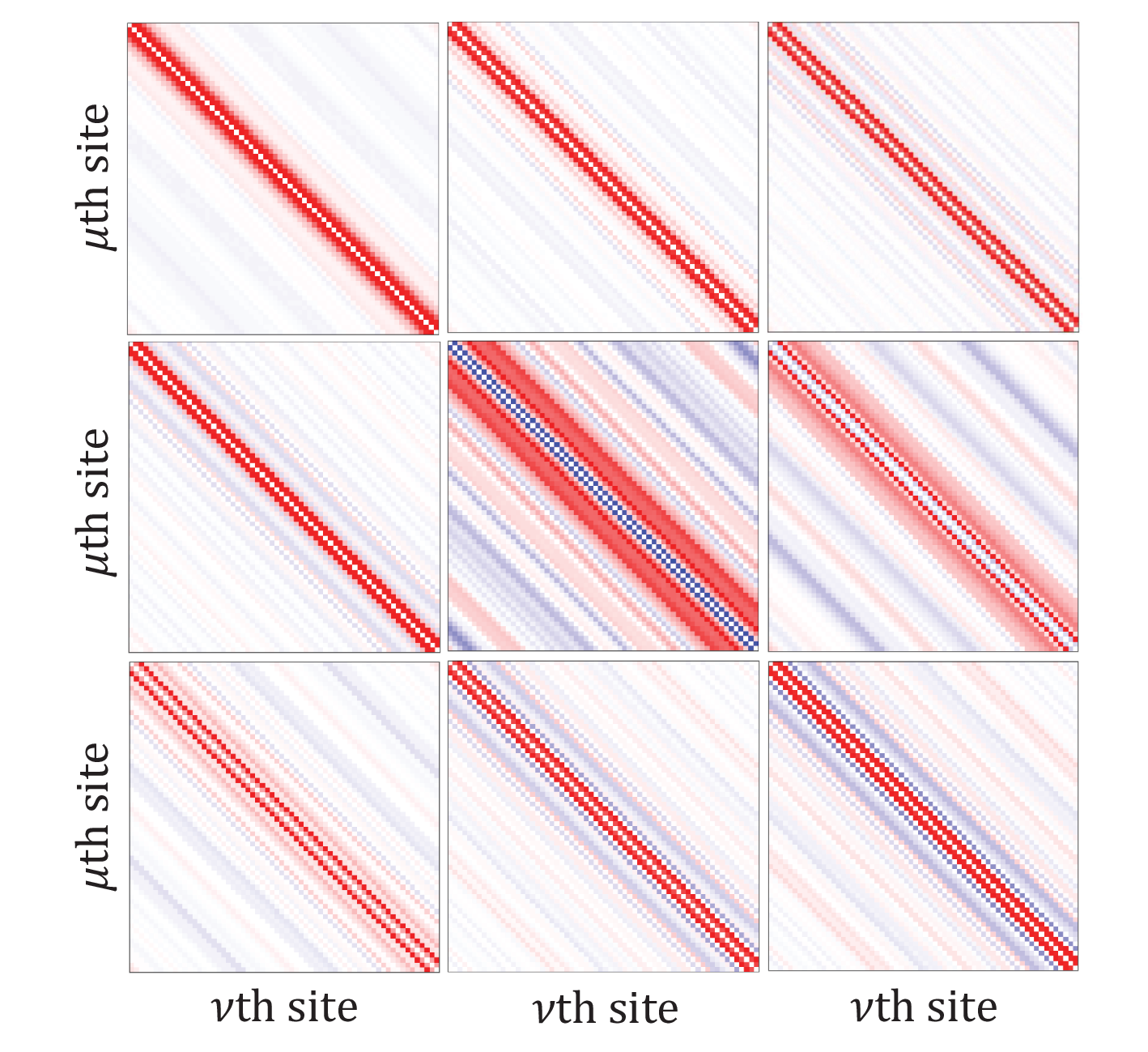}
\caption{Attribution images of crossover from subradiant to superradiant decay rates at $d_s/\lambda= 0.25$ for $N=64$. These colors of images are normalized to their own maximal absolute value of pixels. The indices $m$ of the eigen decay rates $\Gamma_m$ in an ascending order from the upper-left to the lower-right plots are $1$, $9$, $17$, $25$, $33$, $41$, $49$, $57$ and $64$. The eigen decay rate of $m=33$ is the closest one to the natural decay rate $\Gamma$, which shows the contrasting feature of spreading in the off-diagonal parts of attribution images, compared to the importance feature in NN and next NN coupling ranges in the subradiant and superradiant sectors of eigen decay rates.}\label{fig4_s2}
\end{figure}

To differentiate the attributions of the subradiant regime from the superradiant one, we look into the absolute values of attributions in respective regimes. Although they have similar attribution patterns in respective normalized scales of pixels, their absolute values show contrasted scales, where the attribution of the superradiant eigen-decay rates is much larger in orders of magnitudes for small interparticle distances in particular, and thus they can be distinguished. We show the corresponding absolute values of Fig. \ref{fig4} in Fig. \ref{fig4_s1}. For a larger size of the atomic system, we obtain similar attribution features in SSRs and crossover regimes, and as an example, the case of $N=64$ is shown in Fig. \ref{fig4_s2}. For a larger system, alternating features of importance between NN and next NN coupling ranges can appear when approaching the crossover regime from the side of subradiant sector. This shows rich dynamics of attribution features and indicates a possibly hidden rule in the process of correlations build-up in the quantum system.   

\begin{figure}[t]
\centering
\includegraphics[width=8.5cm,height=4cm]{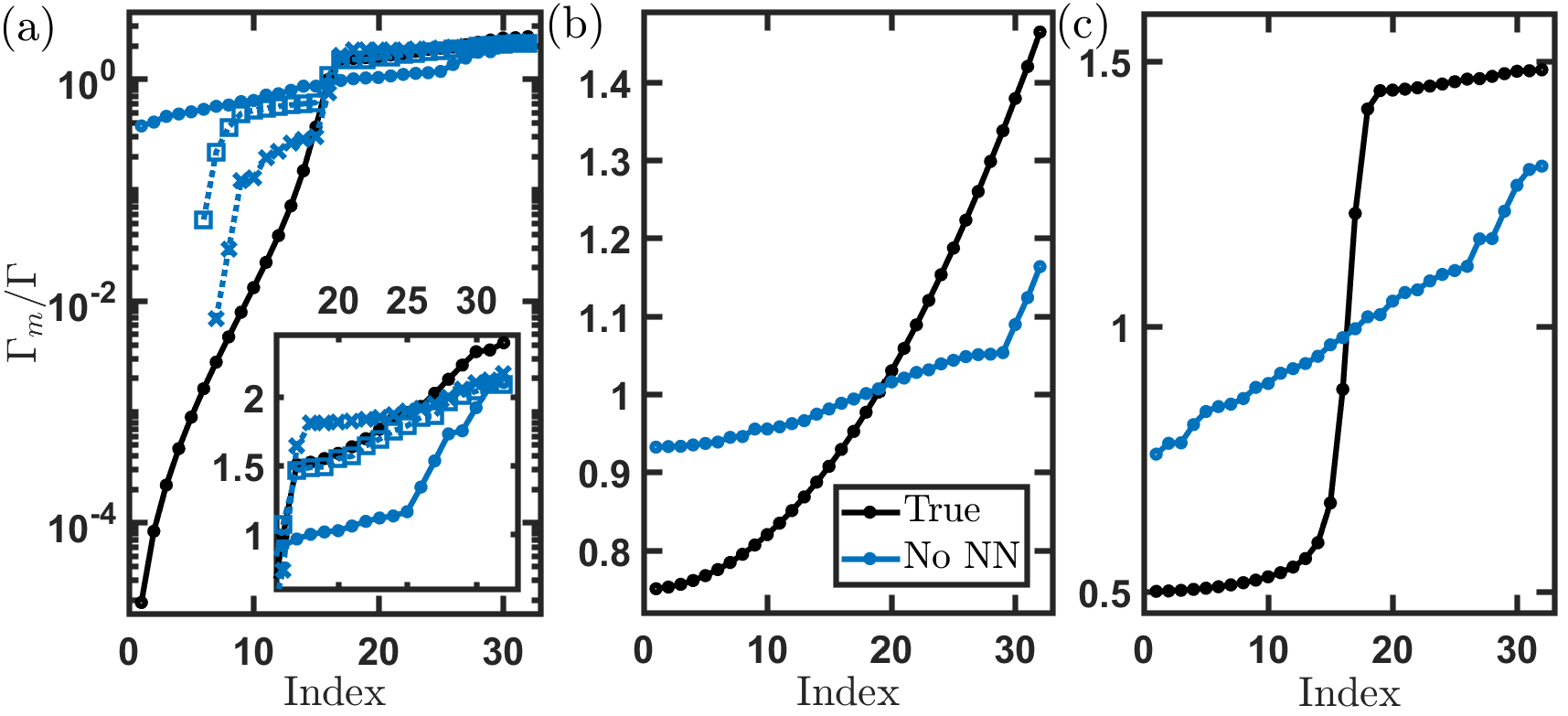}
\caption{Eigen-decay rates $\Gamma_m$ with a removal of NN, next NN, and next next-NN couplings in RDDI for $N=32$. The $\Gamma_m$ are shown in an ascending order in logarithmic scales at (a) $d_s/\lambda=0.25$, and in linear scales at (b) $d_s/\lambda=0.5$ and (c) $d_s/\lambda=0.75$, where true eigen-decay rates with full RDDI are included as a comparison. The deviations of $\Gamma_m$ from the true ones reflect both in the subradiant and superradiant sectors, where an extra removal of next NN ($\times$) or next next-NN coupling term ($\square$) is shown in (a) for a strong interacting regime with an inset of zoom-in plot. In (a), several subradiant decay rates become negative and unphysical, which is similar to Fig. \ref{fig2_tr} owing to the brute-force removals.}\label{fig5}
\end{figure}

As a final remark, we note that the importance feature of RDDI in determining the eigenspectrum can be revealed by designing a partial removal of RDDI elements and identifying its deviation from the true eigenspectrum. We demonstrate this effect in the distributions of eigen-decay rates from the RDDI without NN, next NN, or next next-NN interactions as shown in Fig. \ref{fig5}. This artificial removal leads to significant deviations of the spectrum in both SSRs, where only one particular decay rate near the middle of the spectrum ($\Gamma_m$ with $m\sim N/2$) gets closer to the true one, suggesting that this removal does not make any impact on it. For a strong interacting regime with a smaller interparticle distance in Fig. \ref{fig5}(a), a straight removal of longer-range couplings can lead to unphysical subradiant decay rates, similar in Fig. \ref{fig2_tr} owing to the brute-force removals. To compare the results using attribution technique in Fig. \ref{fig4}, the importance of NN coupling removal coincides with the conclusion drawn in the cases of Figs. \ref{fig4}(a) and \ref{fig4}(c), where significant positive attributions at NN distances show up in the SSRs, while they become negative near the middle of the spectrum. By contrast, the importance of NN coupling is suppressed in Fig. \ref{fig4}(b) in the SSRs. This mismatch may be due to the prevalent feature in the attribution technique we apply here, where all off-diagonal elements of RDDI are considered in contrast to a straight removal of certain part of RDDI elements. For a straight removal of other coupling ranges, the deviation would sustain in SSRs but not near the middle of the spectrum, which indicates a lack of importance feature. The artificial removal of RDDI kernels may provide some clues on how this action modifies the eigenvalue predictions, but a design of the degree of importance requires further exploration. 

\section{Discussion}

It is well known that RDDI in an atomic array are essential in the observations and theoretical predictions of cooperative spontaneous emissions in a finite-size system. However, there is no clear and convincing approach to treat the effect of RDDI in a large and dense atomic ensemble \cite{Andreoli2021} owing to the arising complexity in the emergent and long-range atom-atom correlations. This causes difficulties in numerical simulations by classical computers, and it seems that quantum simulation or quantum computation \cite{DiVincenzo2000} could be the ultimate resolution as they should be for many other complex quantum many-body systems as well. To reveal the essential feature of finite-range correlations induced from RDDI, we take a different route not from physical insights but utilize an interpretable machine as a complimentary instrument. An interpretable machine learning intends to understand the black-box predictions of machine learning \cite{Molnar2020}, with which we are able to obtain the attributions of RDDI from the integrated gradients to connect and relate the predictions of a deep neural network to its input interaction kernels. They provide a causal implication from their features to the structures of RDDI, which can be highly correlated but hidden in the common physical observables of eigenspectrum. 

Our results presented here provide an insightful research direction in quantum systems, which discriminates different sectors of eigenspectrum by distinct features of importance. When the system is under strong RDDI for small interparticle distances, the feature of attributions suggests that at least the next nearest-neighbor coupling range matters for the predictions of the eigen-decay rates. This evidences an insufficient truncation of the coupling range at the nearest-neighbor, and this conclusion supported by the machine learning approach reassures the need for more sophisticated quantum optical theories without resolving the full eigenspectrum to tackle the problem that involves considerable higher-order quantum correlations. For higher-dimensional quantum systems or multiply-excited Hilbert space \cite{Jen2017_MP} under RDDI, we may face new challenges such as extensive quantum correlations emerging from the system, consuming training time in machine learning and the need of new feature importance measures. Aside from this potential difficulty, a higher-dimensional system involve more nearest-neighbor coupling terms in extra spatial dimensions. The results and the conclusion we obtain here for one-dimensional array may not entirely apply to higher-dimensional systems, and long-range interacting feature may arise and become more relevant in the attributions. This needs further clarification and could lead to a future research direction using machine learning approach.     

\begin{figure}[b]
\centering
\includegraphics[width=8.5cm,height=4cm]{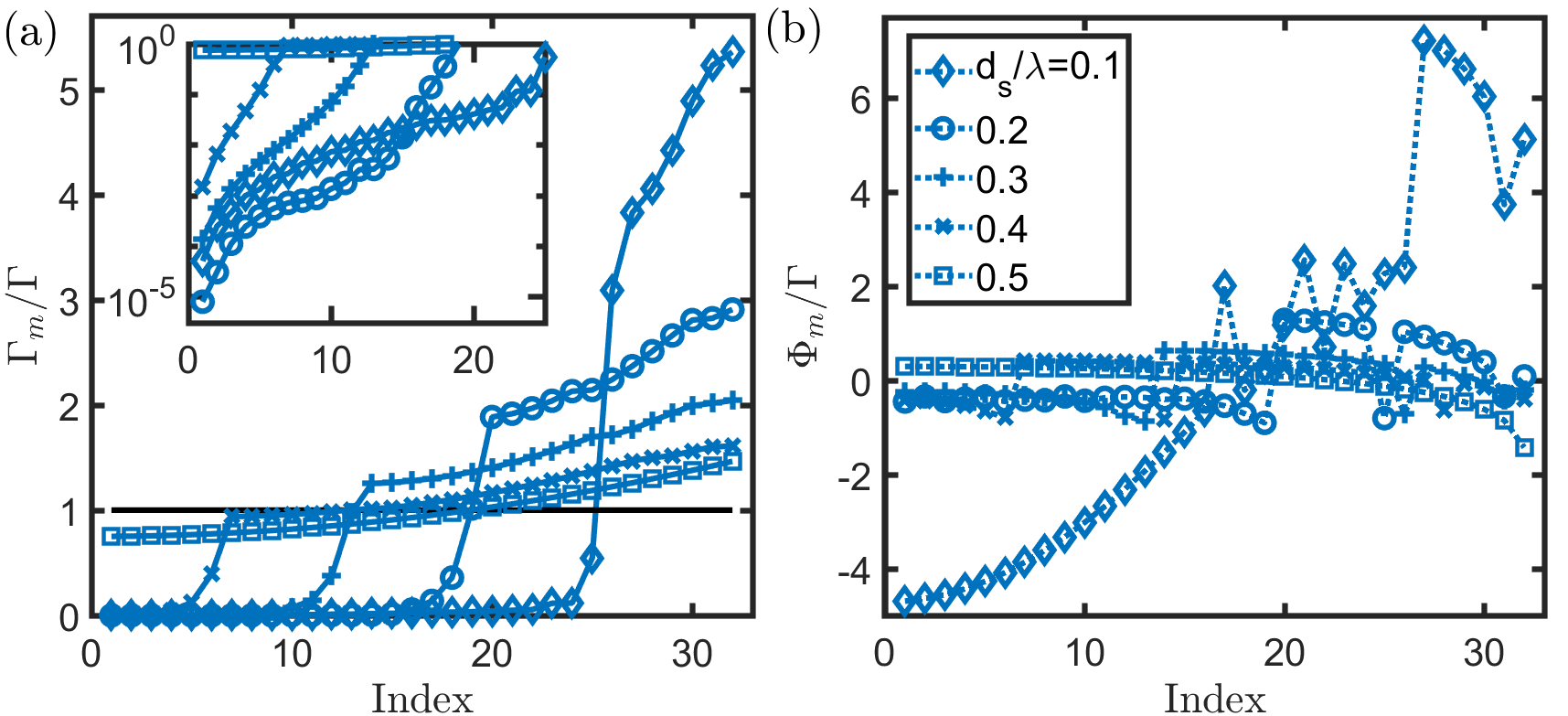}
\caption{Eigen-decay rates $\Gamma_m$ and associated eigen frequencies $\Phi_m$ for $N=32$. The complete eigen-spectrum is plotted for various interparticle distances $d_s$ with $\Gamma_m$ in an ascending order along with corresponding Hermitian counterparts $\Phi_m$.}\label{fig9}
\end{figure}

We note that here we focus on the eigen-decay rates of an equidistant atomic array but neglect their Hermitian counterparts which are eigen frequencies. In general, both decay rates and frequency shifts determine the dynamics of quantum systems. In Fig. \ref{fig9}, we show the results of a full eigen-spectrum for various interparticle distances of $N=32$, where their eigen-decay rates (non-Hermitian parts) along with their Hermitian counterparts $\Phi_m$ are shown in Figs. \ref{fig9}(a) and \ref{fig9}(b), respectively. In the ascending order of eigen-decay rates in Fig. \ref{fig9}(a), we can see the structure and the trend from the subradiant to superradiant sectors. As the distance becomes smaller, strong interacting RDDI leads to a pronounced enhancement of superradiant emissions, and similarly in the subradiant sectors, smaller $d_s$ allows a more subradiant decay behavior, except near $d_s/\lambda=0.25$ where the system presents the most subradiant decay rate (see also Fig. \ref{fig1}). This overall structure of the eigen-decay rates can be observable when multiple superradiant and subradiant eigenmodes are excited, where the scattered photons show an initial abrupt drop, followed by a long-tail radiation \cite{Guerin2016}, signaling the superradiant and subradiant emissions, respectively. As for collective frequency shifts \cite{Scully2009}, they also emerge along with collective radiations, but are harder to be observed in experiments and more sensitive to system’s geometry \cite{Jen2015}. They are often negligible comparing to system’s intrinsic decay rate (up to several hundredths of it in typical cold atoms with a density less than $10^{12}$ cm$^{-3}$) \cite{Pellegrino2014, Meir2014, Bromley2016, Jenkins2016} and can only be significant when a higher density of atomic ensemble is considered ($d_s/\lambda < 0.1$ in average). 

In our considered range of interparticle distances, Fig. \ref{fig9}(b) shows the associated eigen frequencies corresponding to the decay rates in Fig. \ref{fig9}(a). As the interparticle distance increases, eigen frequency becomes less significant as expected. Therefore, the eigen frequency is more related to the strong interacting regime. From the trend and evolution of $\Phi_m$ from the corresponding subradiant to superradiant sectors, we do not see a clear feature associated with the collective decay rates. Moreover, $\Phi_m$ present both positive and negative values in either subradiant or superradiant sector, along with an alternating or oscillatory feature in signs and amplitudes in some regions of the eigen frequencies. This coincides with typical experiments in cold atoms that only little frequency shifts can be observed or theoretically predicted. Nevertheless, we have tried to train the machine to perform simultaneously well in predicting the eigen-decay rates and eigen frequencies, but to no avail. We believe it is this unclear and alternating structure in eigen frequencies that makes our machine difficult to predict the Hermitian counterparts. It would be interesting and challenging as well to resolve the Hermitian part via machine learning method. We also expect new perspectives from machine learning could be drawn in the eigen frequencies, especially for denser atomic systems where long-range quantum correlations emerge and dominate in both subradiant and superradiant sectors of the eigen-spectrum. 

In conclusion, we use the attributions from integrated gradients to explore the features of subradiance and superradiance in the eigenspectrum of an atomic array. The crossover from subradiant to superradiant sectors can be identified in an interpretable machine learning, where the importance of finite-range interaction reflects in the off-diagonal images of attributions, indicating the indispensable role of interaction ranges beyond the nearest-neighbor. In our quantum optical setup with RDDI, a deep network can be used to extract hidden rules from its predictions and provide insights to study other complex many-body condensed-matter systems. 

\section*{ACKNOWLEDGMENTS}
We thank Daw-Wei Wang and his group members in NTHU for insightful discussions on machine learning approach. We also acknowledge support from the Ministry of Science and Technology (MOST), Taiwan, under the Grant No. MOST-109-2112-M-001-035-MY3, and also TensorFlow which provides a free and open-source software library for machine learning.


\end{document}